# Centenary progress from the Nernst theorem to the Nernst statement

Xiaohang Chen[+], Shanhe Su[+], Yinghui Zhou[*], Jincan Chen[*]

Department of Physics, Xiamen University, Xiamen 361005, People's Republic of China

It is found from textbooks that there are the different versions of the schematic diagram related to the Nernst equation, and consequently, it leads to some discussion related to the Nernst equation and the discovery of other meaningful schematic diagrams never appearing in literature. It is also found that through the introduction of a new function, the schematic diagram of the Nernst equation in the isothermal process of any thermodynamic system can be generated in a unified way and that the Nernst equation can be re-obtained from the experimental data of low-temperature chemical reactions without any artificial additional assumptions. The results obtained here show clearly that the centenary progress from the Nernst theorem to the Nernst statement is completed.

[+] The two authors contributed equally to this work.
[*] Emails: yhzhou@xmu.edu.cn; jccchen@xmu.edu.cn

# 1. Introduction

Although the Nernst equation is very important in the theory of thermodynamics [1-7], some artificial additional assumptions were introduced [2, 3, 8, 9]. The Nernst equation implying additional assumptions is referred to as the Nernst theorem and usually used as the core content of the third law of thermodynamics. However, the derivation mode of the Nernst theorem in textbooks has been rarely suspected for over one hundred years. Some questions related to the Nernst equation have not attracted attention until recent [10-13]. The first question is that there are different versions of the schematic diagram related to the Nernst equation in textbooks [2, 3] and literature [14]. Which in those schematic diagrams is correct? The second question is whether there are other meaningful schematic diagrams related to the Nernst equation, which have not yet been given. The third question is whether it is necessary to induction artificial additional assumptions in the setup procedure of the Nernst equation. The fourth question is whether the Nernst equation can be re-obtained from the experimental data of low-temperature chemical reactions without any artificial additional assumptions. These questions are worthy of special discussion.

## 2. The Nernst theorem

In the late 19th century and early 20th century, it was found from an enormous amount of the experimental data at low-temperature chemical reactions that the changes $\Delta H$ and $\Delta G$ of the enthalpy $H$ and Gibbs free energy $G$ of thermodynamic systems in the isothermal and isobaric process become closer and closer as the temperature $T$ decreases. When the temperature is extrapolated to absolute zero [3, 8], there is $(\Delta G)_0 = (\Delta H)_0$, as shown in Fig. 1(a) [3, 9, 15, 16].

In some textbooks [8, 17], it is common to define the chemical affinity $A = -\Delta G$ of a system in the isothermal isobaric process and the heat $Q = -\Delta H$ released by the system. $Q$ is conventionally called the heat of reaction because $Q > 0$ [17] is the more common case for most chemical reactions [8]. By applying the experimental data of $\Delta G$ and $\Delta H$, a schematic diagram of $A$ and $Q$ in the isothermal isobaric process varying with $T$ can be plotted, as shown in Fig. 1 (b) [17,18].

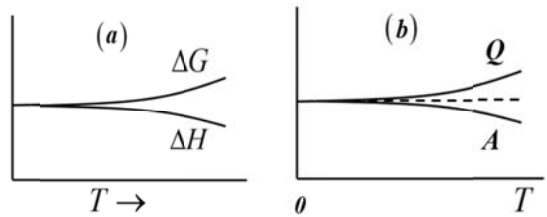

Fig.1. The conventional schematic diagrams of (a) $\Delta H$ and $\Delta G$ and (b) $Q$ and $A$ of the thermodynamic systems in the isothermal and isobaric process varying with $T$.

According to the definition of the Gibbs free energy

$$G = H - TS, \qquad (1)$$

the fundamental relation between $\Delta H$ and $\Delta G$ of a thermodynamic system in the isothermal process is given by

$$\Delta G = \Delta H - T\Delta S, \qquad (2)$$

where $\Delta S$ is the entropy change of a thermodynamic system in an isothermal process. Using two artificial additional assumptions typically introduced in textbooks [3, 9], i.e., $\Delta S$ is bounded when $T \to 0$ and $\lim_{T \to 0}(\partial \Delta H / \partial T) = \lim_{T \to 0}(\partial \Delta G / \partial T)$, one can derive the Nernst equation, i.e.,

$$\lim_{T \to 0}(\Delta S)_T = 0. \tag{3}$$

Because the additional assumptions are introduced in the derivative process, the Nernst equation is also called the Nernst postulate or Nernst theorem [3, 9, 19], which has been taken as the core contents of the third law of thermodynamics. It shows clearly that the Nernst equation is very important in the theory of thermodynamics [1-6]. However, the four questions raised in Introduction are closely related to the Nernst equation and rarely mentioned [20]. Practically, these questions are very worthy of special discussion.

**3. The schematic diagrams related to the Nernst theorem**

Considering the fact that $\Delta H$ and $\Delta G$ are the experimental data obtained from the thermodynamic systems in the isothermal and isobaric process, one can determine $\Delta G < 0$ when $T > 0$, because the irreversible process that occurs in the isothermal and isobaric system always proceeds in the direction of the reduction of the Gibbs free energy [9, 21]. There are two cases for $\Delta H$, i.e., $\Delta H > 0$ and $\Delta H < 0$, which will be, respectively, discussed below.

(i) The case of $\Delta H > 0$. In the region of $T > 0$, $\Delta H > 0 > \Delta G$. When the experimental data of low temperature chemical reactions are extrapolated to absolute zero, there is $(\Delta G)_0 = (\Delta H)_0$. Based on the two conditions mentioned just, it is necessary to have $(\Delta H)_0 = 0$. Thus, the curves of $\Delta H$ and $\Delta G$ varying with $T$ should be schematically shown in Fig. 2 (a).

(ii) The case of $\Delta H < 0$. In the temperature range of $T > 0$, $\Delta H = \Delta Q < 0$ for a simple thermodynamic system in the isothermal and isobaric process, where $\Delta Q$ indicates the heat absorbed by the system. $\Delta Q < 0$ represents that the system releases heat to the environment in the isothermal isobaric process. For a reversible isothermal exothermic process, $\Delta S < 0$ can be directly obtained from $\Delta Q = T\Delta S < 0$. For an irreversible isothermal exothermic process, $\Delta Q < T\Delta S$. It cannot be judged from both $\Delta Q < 0$ and $\Delta Q < T\Delta S$ whether $\Delta S$ is smaller than 0 or not. The entropy change $\Delta S$ of the system in an irreversible isothermal exothermic process is composed of the entropy decrease caused by the heat released in the isothermal process and the entropy increase caused by the irreversibility inside the system. $\Delta S < 0$ holds only when the entropy change caused by isothermal heat release is dominant. It can be seen from Eq. (2) that in such a case, $0 > \Delta G > \Delta H$. When $T \to 0$, $(\Delta G)_0 = (\Delta H)_0$. The curves depicting the variations of $\Delta G$ and $\Delta H$ with respect to $T$ should be schematically illustrated in Fig. 2(b). When the entropy change caused by the irreversibility inside the system is dominant, $\Delta S > 0$. It be seen from Eq. (2) that $0 > \Delta H > \Delta G$. When $T \to 0$, $(\Delta G)_0 = (\Delta H)_0$. The curves of $\Delta G$ and $\Delta H$ varying with $T$ should be schematically illustrated in Fig. 2(c). When the entropy change caused by the isothermal heat release is equal to that caused by the irreversibility inside the system, $\Delta S = 0$, as shown in Fig. 2(d). This is exactly the case what textbook [3] describes: the enthalpy change $\Delta H$ is very nearly equal to the Gibbs free energy change $\Delta G$ over a considerable temperature range.

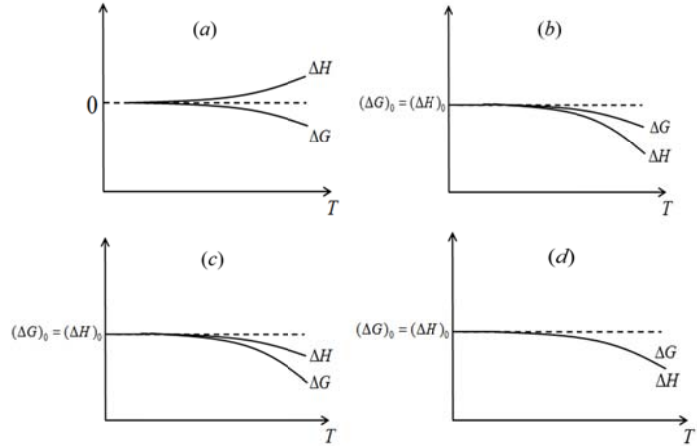

Fig. 2. The schematic diagrams of $\Delta H$ and $\Delta G$ of the thermodynamic systems in the isothermal and isobaric process varying with $T$, where (a) $\Delta H > 0 > \Delta G$, (b) $0 > \Delta G > \Delta H$, (c) and (d) $0 > \Delta H = \Delta G$. When $T \to 0$, (a) $(\Delta G)_0 = (\Delta H)_0 = 0$ and (b)-(d) $(\Delta G)_0 = (\Delta H)_0$.

It can be clearly seen from Eq. (1) that the isothermal isobaric process mentioned above is only a sufficient condition that Eq. (2) is true, while the isothermal process is a necessary condition that Eq. (2) is true. It shows clearly that in low temperature experiments, the test can be carried out in the isothermal isobaric process as well as in the isothermal isovolumetric or isothermal process. Now, we continue to discuss that the cases of $\Delta H$ and $\Delta G$ of the thermodynamic systems in the isothermal isovolumetric or isothermal process varying with $T$.

For a simple thermodynamic system only including volume variable work, the enthalpy change in the isothermal isovolumetric process can be expressed as

$$\Delta H = \Delta Q + v\Delta P, \qquad (4)$$

where the change of the Gibbs free energy is still represented by Eq. (2), and $v$ and $P$ are the volume and pressure of the system. When the isothermal isovolumetric process is irreversible, $\Delta Q < T\Delta S$. According to Eqs. (2) and (4), $\Delta G < v\Delta P$. In the region of $T > 0$, when $\Delta H > v\Delta P$, there is $\Delta H > v\Delta P > \Delta G$; When $T \to 0$, $(\Delta H)_0 = (\Delta G)_0 = (v\Delta P)_0$; as shown in Fig. 3 (a). When $\Delta H < v\Delta P$, $\Delta Q < 0$ may be determined by Eq. (4). This means that the system releases heat to the environment. In the region of $T > 0$, $\Delta Q < 0$ and $\Delta Q < T\Delta S$ cannot determine whether $\Delta S$ is less than zero. As described above, there are three cases for $\Delta S$. When $\Delta S < 0$, $(v\Delta P) > \Delta G > \Delta H$; When $\Delta S > 0$, $(v\Delta P) > \Delta H > \Delta G$; When $\Delta S = 0$, $(v\Delta P) > \Delta G = \Delta H$; as shown in Figs. 3 (b) - (d), respectively. When $T \to 0$, $(\Delta H)_0 = (\Delta G)_0 = (\Delta Q)_0 + (v\Delta P)_0$.

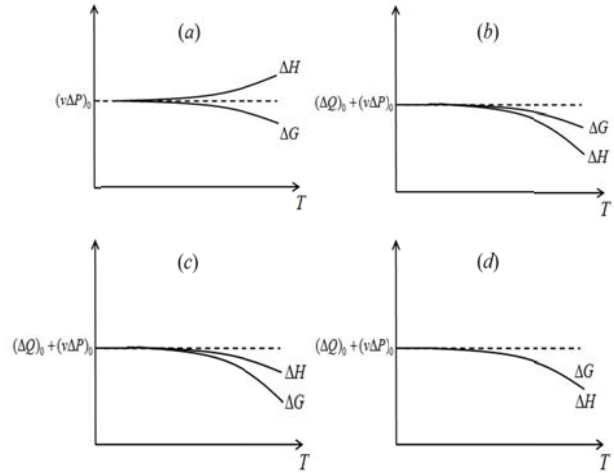

Fig. 3. The schematic diagrams of $\Delta H$ and $\Delta G$ of the thermodynamic systems in the isothermal process varying with $T$, (a) $\Delta H > (v\Delta P) > \Delta G$, (b) $(v\Delta P) > \Delta G > \Delta H$, (c) $(v\Delta P) > \Delta H > \Delta G$, and (d) $(v\Delta P) > \Delta H = \Delta G$. When $T \to 0$, （a） $(\Delta H)_0 = (\Delta G)_0 = (v\Delta P)_0$, and （b）-（d） $(\Delta H)_0 = (\Delta G)_0 = (\Delta Q)_0 + (v\Delta P)_0$.

For the isothermal process, the small changes of the volume $v$ and pressure $P$ of the system are, respectively, $\Delta v$ and $\Delta P$, and the internal energy and enthalpy changes can be expressed as

$$\Delta U = \Delta Q - (P + \Delta P/2)\Delta v \qquad (5)$$

and

$$\Delta H = \Delta Q + (v - \Delta v/2)\Delta P. \qquad (6)$$

Setting $v' = v - \Delta v/2$ and substituting $v'$ for $v$ in Fig. 3, we can directly obtain the schematic diagrams of $\Delta H$ and $\Delta G$ of the thermodynamic systems in the irreversible isothermal process varying with $T$. If the higher order small quantity $\Delta P \Delta v/2$ is ignored, Eq. (6) is equal to Eq. (4) and the schematic diagrams of $\Delta H$ and $\Delta G$ of the thermodynamic systems in the irreversible isothermal process varying with $T$ are the same as those shown by Fig.3.

According to Figs. 2 and 3 and the definitions of $A$ and $Q$, one can easily generate the curves of $Q$ and $A$ varying with $T$. For example, the schematic diagrams of $Q$ and $A$ of the thermodynamic systems in the isothermal and isobaric process varying with $T$ are shown in Fig. 4. It is meaningful to note that one schematic diagram similar to Fig.4 (a) had been adopted in the early textbook [22], but the value of $Q$ at absolute zero temperature was not determined.

According to Figs. 2-4, Eq. (2), and two artificial additional assumptions mentioned above, one can conveniently derive Eq. (3).

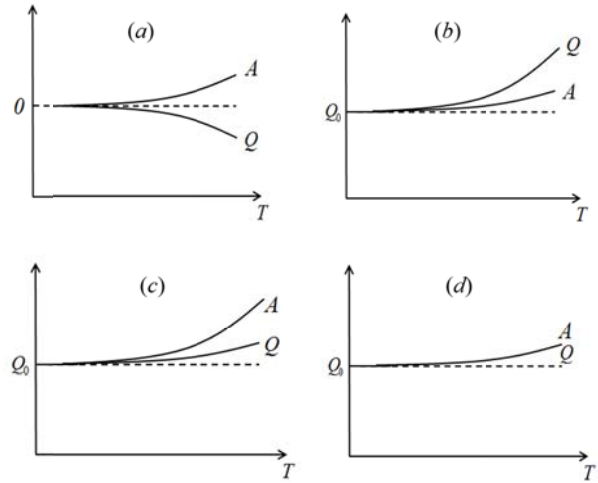

Fig.4. The schematic diagrams of $Q$ and $A$ of the thermodynamic systems in the isothermal and isobaric process varying with $T$, where (a) $A > 0 > Q$, (b) $Q > A > 0$, (c) $A > Q > 0$, and (d) $A = Q > 0$. When $T \to 0$ (a) $A_0 = Q_0 = 0$ and (b)-(d) $A_0 = Q_0$.

It is seen clearly from the Nernst equation that $\Delta S$ is the entropy change of a thermodynamic system in an isothermal process and may be obtained from the experimental data of different parameters of thermodynamic systems at low-temperatures. For example, $\Delta S$ can be obtained from $\Delta H$ and $\Delta G$ as well as $\Delta U$ and $\Delta F$ of the thermodynamic systems at low-temperatures, where $U$ and $F$ are the internal energy and free energy of a thermodynamic system, respectively.

According to the definition of the free energy

$$F = U - TS, \tag{7}$$

the fundamental relation between $\Delta U$ and $\Delta F$ of a thermodynamic system in the isothermal process is given by

$$\Delta F = \Delta U - T\Delta S. \tag{8}$$

For a simple thermodynamic system only including volume variable work, the internal energy change and the free energy change in the isothermal isovolumetric process can be, respectively, expressed as

$$\Delta U = \Delta Q \leq T \Delta S \tag{9}$$

and

$$\Delta F \leq 0. \tag{10}$$

Considering the fact that $\Delta U$ and $\Delta F$ are the experimental data obtained from the thermodynamic systems in the irreversible isothermal isovolumetric process, one can determine $\Delta F < 0$ when $T > 0$, because the irreversible process that occurs in the isothermal isovolumetric system always proceeds in the direction of the reduction of the free energy. There are two cases for $\Delta U$, i.e., $\Delta U > 0$ and $\Delta U < 0$, which will be, respectively, discussed below.

(i) The case of $\Delta U > 0$. In the region of $T > 0$, $\Delta U > 0 > \Delta F$. When the experimental data of low temperature chemical reactions are extrapolated to absolute zero, there should be $(\Delta F)_0 = (\Delta U)_0$ and it is necessary to have $(\Delta U)_0 = 0$. Thus, the curves of $\Delta U$ and $\Delta F$ varying with $T$ should be schematically shown in Fig. 5 (a). It is found that the schematic diagram similar to Fig. 5(a) have appeared in textbook [23] for several times. In textbook [23], $U$ and $A$ indicate $\Delta U$ and $\Delta F$, respectively. This shows that the experimental data of $\Delta U$ and $\Delta F$ of thermodynamic systems at low temperatures can be used to derive the Nernst equation.

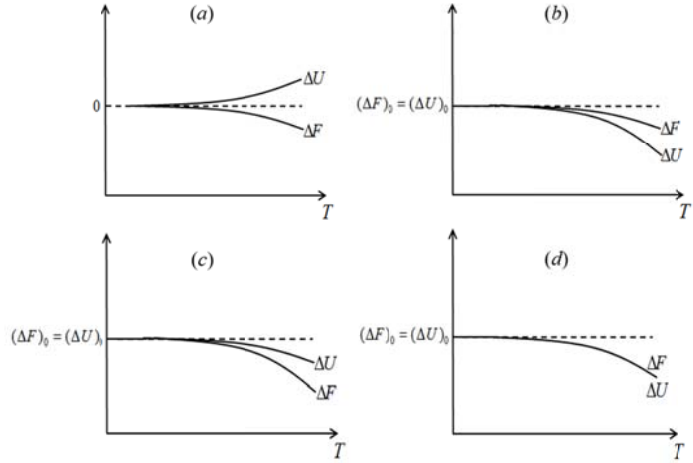

Fig. 5. The schematic diagrams of $\Delta U$ and $\Delta F$ of the thermodynamic systems in the isothermal isovolumetric process varying with $T$, where (a) $\Delta U > 0 > \Delta F$, (b) $0 > \Delta F > \Delta U$, (c) $0 > \Delta U > \Delta F$, and (d) $0 > \Delta U = \Delta F$. When $T \to 0$, (a) $(\Delta F)_0 = (\Delta U)_0 = 0$ and (b)-(d) $(\Delta F)_0 = (\Delta U)_0$.

(ii) The case of $\Delta U < 0$. In the temperature range of $T > 0$, $\Delta U = \Delta Q < 0$ for a simple thermodynamic system in the isothermal isovolumetric process, which indicates that the system releases heat to the environment. For an irreversible isothermal exothermic process, $\Delta Q < T\Delta S$. It cannot be judged from both $\Delta Q < 0$ and $\Delta Q < T\Delta S$ whether $\Delta S$ is smaller than 0 or not. As described above, there are three cases for $\Delta S$. (i) $\Delta S < 0$ and $0 > \Delta F > \Delta U$. (ii) $\Delta S > 0$ and $0 > \Delta U > \Delta F$. (iii) $\Delta S = 0$ and $0 > \Delta F = \Delta U$. When $T \to 0$, $(\Delta F)_0 = (\Delta U)_0$. The curves depicting the variations of $\Delta F$ and $\Delta U$ with respect to $T$ should be schematically illustrated in Fig. 5(b)-(d).

For a simple thermodynamic system only including volume variable work, the internal change in the isothermal isobaric process can be expressed as

$$\Delta U = \Delta Q - P\Delta v, \tag{11}$$

where the change of the free energy is still represented by Eq. (8). According to Eqs. (8) and (11), $\Delta F < -P\Delta v$ for an irreversible isothermal isobaric process In the region of $T > 0$, when $\Delta U > -P\Delta v$, there is $\Delta U > -P\Delta v > \Delta F$; When $T \to 0$, $(\Delta U)_0 = (\Delta F)_0 = (-P\Delta v)_0$; as shown in Fig. 6 (a). When $\Delta U < -P\Delta v$, $\Delta Q < 0$ may be determined by Eq. (11). This means that the system releases heat to the environment. In the region of $T > 0$, $\Delta Q < 0$ and $\Delta Q < T\Delta S$ cannot determine whether $\Delta S$ is less than zero or not. As described above, there are three cases for $\Delta S$. When $\Delta S < 0$, $(-P\Delta v) > \Delta F > \Delta U$; When $\Delta S > 0$, $(-P\Delta v) > \Delta U > \Delta F$; When $\Delta S = 0$, $(-P\Delta v) > \Delta F = \Delta U$; as shown in Figs. 6 (b) - (d), respectively. When $T \to 0$, $(\Delta U)_0 = (\Delta F)_0 = (\Delta Q)_0 - (P\Delta v)_0$.

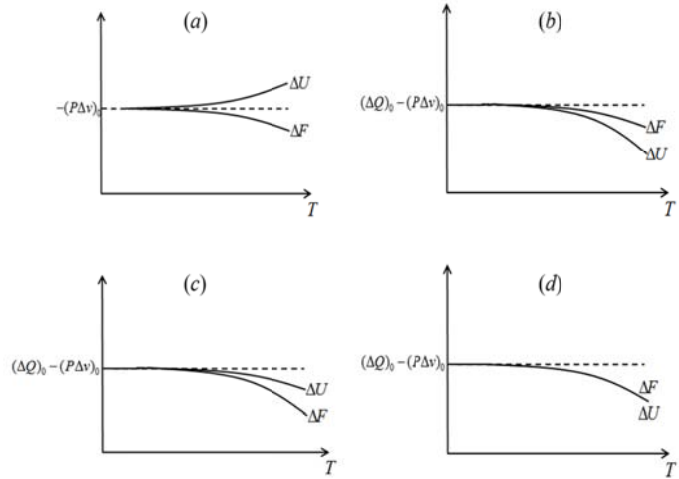

Fig. 6. The schematic diagrams of $\Delta U$ and $\Delta F$ of the thermodynamic systems in the isothermal process varying with $T$, (a) $\Delta U > -P\Delta v > \Delta F$, (b) $(-P\Delta v) > \Delta F > \Delta U$, (c) $(-P\Delta v) > \Delta U > \Delta F$, and (d) $(-P\Delta v) > \Delta F = \Delta U$. When $T \to 0$, (a) $(\Delta U)_0 = (\Delta F)_0 = (-P\Delta v)_0$, and (b) - (d) $(\Delta U)_0 = (\Delta F)_0 = (\Delta Q)_0 - (P\Delta v)_0$.

For an isothermal process with small changes $\Delta v$ and $\Delta P$, the internal energy change and the free energy change are still given by Eqs. (5) and (8). Setting $P' = P + \Delta P/2$ and substituting $P'$ for $P$ in Fig. 6, we can directly obtain the schematic diagrams of $\Delta U$ and $\Delta F$ of the thermodynamic systems in the irreversible isothermal process varying with $T$. If the higher order small quantity $\Delta P \Delta v/2$ is ignored, Eq. (5) is equal to Eq. (11) and the schematic diagrams of $\Delta U$ and $\Delta F$ of the thermodynamic systems in the irreversible isothermal process varying with $T$ are the same as those shown by Fig. 6.

According to Figs. 5 and 6, Eq. (8), and two similar additional assumptions mentioned above, i.e., $\Delta S$ is bounded when $T \to 0$ and $\lim_{T \to 0}(\partial \Delta U / \partial T) = \lim_{T \to 0}(\partial \Delta F / \partial T)$, one can conveniently derive Eq. (3), i.e., the Nernst equation.

For the isothermal process of a thermodynamic system, $\Delta Q \leq T \Delta S$. When the system is surrounded by an environment much larger than it, both of which are at absolute zero temperature, $\Delta Q \leq 0$ as long as Eq. (3) holds. This means that during the isothermal process of $T = 0$, endothermic heat does not occur. Similarly, the heat absorption of the environment at $T = 0$ will also not occur. If the isothermal process of the thermodynamic system is irreversible, the system must release heat, but the environment will not absorb heat, so that the irreversible isothermal process of the thermodynamic system cannot be carried out, as shown in Fig.7. This shows that the isothermal process of a thermodynamic system at $T = 0$ must be reversible as long as Eq. (3) holds. As described in textbooks, the isothermal process carried out by the thermodynamic system at $T=0$ coincides with the reversible adiabatic process [3, 9], and there is no heat exchange between the system and the environment. Thus, $(\Delta H)_0$ in Figs. 2(b)-(d) and $(\Delta U)_0$ in Figs. 5(b)-(d) are equal to zero, and $(\Delta Q)_0$ in Figs. 3(b)-(d) and 6(b)-(d) and $Q_0$ in Figs. 4(b)-(d) are also equal to zero. However, $(v\Delta P)_0$ in Fig.3 for the thermodynamic systems that is not in the isothermal isobaric process and $(P\Delta v)_0$ in Fig.6 for the thermodynamic systems that is not in the isothermal isovolumetric process are not required to equal to zero so that $(\Delta H)_0$ and $(\Delta U)_0$ may not be equal to zero. These energy $(v\Delta P)_0$ and $(P\Delta v)_0$ translate as the internal energy of the system. When $(v\Delta P)_0 < 0$ or $(P\Delta v)_0 > 0$ the system does work on the environment.

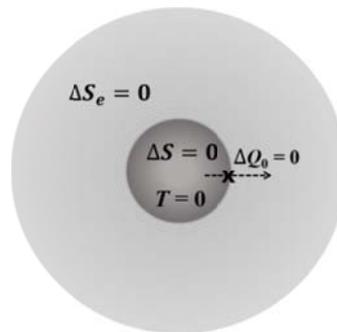

Fig.7. The schematic diagram of a thermodynamics system at $T=0$.

It can be found to compare Figs.2 and 4 with Fig.1 that Figs. 1(a) and (b) are incorrect, because the schematic diagrams in Fig.1 do not satisfy the basic requirement of $\Delta H > 0 > \Delta G$ or ($0 > \Delta G, 0 > \Delta H$) in the region of $T > 0$. The errors in Fig. 1 may originate in Ref. [18] or in other previous references. Unfortunately, these wrong diagrams have been ignored for a long time and adopted by most textbooks [3, 9, 14-17] for more than one century. It can be also found that Figs. 2 (b)-(d), 4 (b)-(d), and 5(b)-(d) never appear in textbooks and literature and are the new schematic diagrams related to the Nernst equation, which should be paid attention to in the teaching and textbook compilation of universities. Figs.3 and 6 may be some meaning schematic diagrams related to the Nernst equation, but these diagrams have never been seen in history. So far we have answered the first and second questions raised in Introduction.

**4. The Nernst statement**

It can be seen from Figs. 2-6 that when the curves of $\Delta H$ and $\Delta G$ (or $\Delta U$ and $\Delta F$) of a thermodynamic system varying with $T$ are plotted, it is necessary to indicate whether the system is in an isothermal isobaric process, an isothermal isovolumetric process, or an isothermal process. It shows clearly that even for a same thermodynamic system, the schematic diagrams of the Nernst equation may be different for different experimental test conditions. It can be also seen from Figs. 2-6 that these schematic diagrams are suitable for the thermodynamic systems only including volume variable work. It means that for different thermodynamic systems, the schematic diagrams of the Nernst equation are different. In particular, for general thermodynamic systems described by the following equation [3, 24-26]

$$dU = TdS + \sum_{i=1}^{n} Y_i dy_i, \tag{12}$$

it is highly meaningful to know how to generate the schematic diagrams of the Nernst equation. In Eq. (12), $y_i$ and $Y_i$ are the generalized coordinates and corresponding generalized forces [10, 12, 27], and $n$ is the number of generalized coordinates.

It was found from an amount of experimental data at low temperatures that $\Delta H$ ($\Delta U$) of a thermodynamic system at different experimental test conditions are very different and so are $\Delta G$ ($\Delta F$). However, they have a common feature: no matter whether the thermodynamic system is a simple system or a general system and whether the system is in an isothermal isobaric process, an isothermal isovolumetric process, or an isothermal process, the experimental data of $\Delta H$ and $\Delta G$ (or $\Delta U$ and $\Delta F$) become closer and closer as the temperature decreases. When $T \rightarrow 0$, $(\Delta G)_0 = (\Delta H)_0$ and $(\Delta F)_0 = (\Delta U)_0$. With the help of this feature, one can innovatively introduce a new function

$$f \equiv |\Delta H - \Delta G| = |\Delta U - \Delta F| \tag{13}$$

and easily generate the curves of $f$ varying with $T$, as indicated in Fig.8. The importance of Fig.8 lies in the fact that the experimental data of $\Delta H$ and $\Delta G$ as well as $\Delta U$ and $\Delta F$ may be adopted. It shows that Fig.8 is suitable for not only a simple thermodynamic system but also a general thermodynamic system [12].

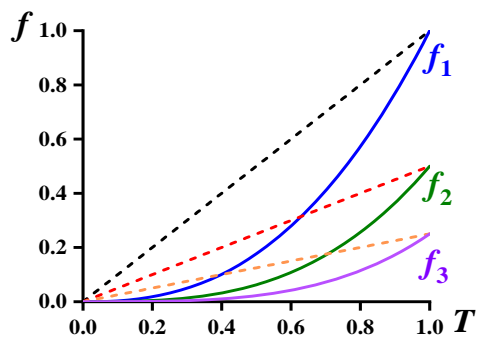

Fig. 9. The schematic diagram of the new function $f$ of the thermodynamic systems in the isothermal process varying with $T$.

It is seen from Fig.8 that $f_i$ ($i = 1, 2, ...$) is a function of $T$. For different thermodynamic systems, $f_i$ may have different forms. It is found from a large number of experimental data that for general thermodynamic systems at ultra-low temperatures, $f_i$ has one common feature: The decrease of $f_i$ with temperature is faster than that of $T$ itself [12, 13, 28]. The general expression of $f_i$ may be given by

$$f_i = a_{i1}T^\alpha + a_{i2}T^\beta + a_{i3}T^\gamma + ..., \tag{14}$$

where $a_{ij} \geq 0$ ($j = 1, 2, 3, ...$) are some coefficients to be independent of temperature, and $1 < \alpha < \beta < \gamma$. The simplest form of $f_i$ is $f_i = a_{i1}T^\alpha$.

It is seen from the curves of $f_i$ varying with $T$ that the lower the temperature is, the smaller the slope of curves, resulting in a result that the absolute value of the entropy change $|(\Delta S)_T|$ of thermodynamic systems during the isothermal process becomes smaller. When $T \to 0$, the slope of curves approaches zero and $(\Delta S)_T = 0$. Moreover, substituting $f_i$ into Eqs. (2) and (8) and extrapolating it to absolute zero temperature, one can determine

$$\lim_{T \to 0} \frac{|\Delta H - \Delta G|}{T} = \lim_{T \to 0} \frac{|\Delta U - \Delta F|}{T} = \lim_{T \to 0} \frac{f_i}{T} = \lim_{T \to 0} |(\Delta S)_T| = 0 = \lim_{T \to 0} (\Delta S)_T. \tag{15}$$

The last equation in Eq. (15) is exactly the Nernst equation, i.e., Eq. (3). It should be emphasized here that without any artificial additional assumptions, the Nernst equation can be directly obtained from the experimental data of thermodynamic systems at low-temperature chemical reactions. This is obviously different from the methods in textbooks, where some artificial additional assumptions were introduced. Thus, the physical contents included in the Nernst equation should not be referred to as the Nernst postulate or the Nernst theorem [1-4, 29-34] in textbooks and literature. The relevant contents of the Nernst equation in textbooks should be rewritten. The conventional terminologies, such as the Nernst postulate or the Nernst theorem, prevalent in textbooks and literature, should be expunged and aptly renamed as the Nernst statement. Hitherto, we have solved the third and fourth problem raised in Introduction

There are usually two different views on the Nernst equation in textbooks. The first view [8, 9, 35] is that the entropy change associated with any isothermal process of a condensed system approaches zero as the temperature approaches absolute zero, and the second view [3, 4, 17] is that the entropy change associated with any isothermal reversible process of a condensed system approaches zero as the temperature approaches absolute zero. It can be clearly seen from Figs. 2-6 and 8 that the experimental measurements are usually completed in the irreversible process in the region of $T > 0$. When the temperature is extrapolated to absolute zero, the isothermal isobaric process, isothermal isovolumetric process, and isothermal process in Figs. 2-6 and 8 tend to be reversible. Thus, the second view is more accurate than the first and is acceptable. However, it does not contain the characteristics of irreversible isothermal processes and entropy changes of the thermodynamic system in the region of $T > 0$. Therefore, the physical connotation contained in the Nernst equation can be interpreted as the absolute value of the entropy change of any thermodynamic system in the isothermal process decreases with the decrease of temperature; when the absolute temperature tends to zero, the isothermal process tends to be reversible and the entropy change of the system approaches zero. Such a statement may be called the Nernst statement and is more comprehensive and accurate than those given in textbooks, not only including the performance changes of thermodynamic systems at low temperatures but also revealing the characteristics of thermodynamic systems at $T = 0$.

## 5．Conclusions

The Nernst equation can be re-obtained from the experimental data of low-temperature chemical reactions without any artificial additional assumptions. This realizes the progress from the Nernst theorem to the Nernst statement. From now on, the Nernst statement will replace the Nernst theorem as one statement of the third law of thermodynamics, which can solve the awkward problem caused by the Nernst theorem used as the core content of the third law of thermodynamics, so that the third law of thermodynamics is a true reflection of the objective world.